\documentstyle[12pt]{article}
\begin{document}
\setlength{\unitlength}{1mm}

{\hfill Preprint DSF-13/94}\\

{\hfill May 1994}\\
\begin{center}
{\Large\bf Spectral Geometry and One-loop Divergences}\\
\end{center}
\begin{center}
{\Large\bf on Manifolds with Conical Singularities}
\end{center}

\bigskip\bigskip

\begin{center}
{\bf Dmitri V. Fursaev}\footnote{e-mail: fursaev@na.infn.it and
fursaev@theor.jinrc.dubna.su after August 1.}
\end{center}

\begin{center}
{{\it Laboratory of Theoretical Physics, Joint Institute for
Nuclear Research, \\
Head Post Office, P.O.Box 79, Moscow, Russia\\
and\\
Dipartimento di Scienze Fisiche, Universit\`a di Napoli -
Federico II -, and INFN\\
Sezione di Napoli, Mostra D'Oltremare Pad. 19, 80125, Napoli, Italy}\\}
\end{center}

\bigskip\bigskip\bigskip

\begin{abstract}
Geometrical form of the one-loop divergences induced by
conical singularities of background manifolds is studied.
To this aim the heat kernel asymptotic expansion
on spaces having the structure $C_{\alpha}\times \Sigma$
near singular surface $\Sigma$ is analysed.
Surface corrections to standard second and third heat coefficients
are obtained explicitly in terms of angle $\alpha$ of a  cone
$C_{\alpha}$ and components of the Riemann tensor.
These results are compared to ones
to be already known for some particular cases.
Physical aspects of the surface divergences are
shortly discussed.
\end{abstract}

\vspace{7cm}

\newpage
\baselineskip=.8cm

\section{Introduction}

There is a number of physical problems where quantum
field effects in presence of conical singularities of the
background manifolds play an important role.
For instance, such singularities are induced by
idealized cosmic strings and in this case the conical angle deficit
is determined by the string tension parameter \cite{a4}. It gives rise to
some interesting quantum phenomena like
vacuum polarization around strings \cite{a17}, \cite{a13} and to
necessity in renormalization of the string tension \cite{b5}.

Finite temperature
quantum field theory on static spaces with bifurcate Killing horizons
represents another example.
Typically these spaces have topology $R^2\times S^2$, where
$S^2$ is a bifurcation surface that is unchanged under action
of the isometry group. Introduction of thermal
equilibrium here is equivalent in general to passing to singular
Euclidean space-times
with topology $C_{\alpha}\times S^2$ where conical angle $\alpha$ of
$C_{\alpha}$ is
associated to the inverse temperature. Cosmological implications
of this theory in de Sitter space-time have been considered in \cite{a6},
\cite{b1}. On the other hand, the importance of such singular geometry
has been recently pointed out in connection with statistical mechanical
computations of the black hole entropy \cite{b2}, \cite{b4}, when
the temperature different from the Hawking one should be introduced
to get the corresponding derivative of the partition function.
In particular, it results to additional ultraviolet divergent terms
in the Bekenstein-Hawking entropy. However, an explicit geometrical structure
of
these terms, needed to understand how to handle the new divergences,
is not yet known.

In present letter we find the structure of the one-loop divergences
induced by such conical defects.
For this aim the asymptotic expansion of the
trace of the heat operator on a compact manifold $M_{\alpha}$ with
singularities having the form
$C_{\alpha}\times \Sigma$, where $\Sigma$ is the surface of fixed points
of one-parameter isometry group of $M_{\alpha}$, is analysed and the first
three heat coefficients are obtained explicitly.
Our main result
is formulated in section 2. Its proof and
comparison with some particular cases are given in section 3.
Then, we shortly discuss physical aspects of addidional
surface counterterms in the effective action and present our conclusions.

\section{The result}

The most convenient way to investigate the structure of one-loop
divergences of
the effective action is to express the contributions of
all one-loop diagrams, that can be sum up in a determinant, through
the trace of the heat kernel operator. In particular, for
logarithm of the scalar field determinant to appear in the one-loop
effective action one can use the DeWitt-Schwinger proper time
representation \cite{a16}
\begin{equation}
\log\det(-\Box + \xi R + m^2)=-\int_{0}^{\infty}{ds \over s}Tr K_M(s) e^{-m^2s}
\label{eq:log}
\end{equation}
where $\Box =g^{-1/2}\partial _{\mu}g^{1/2}g^{\mu\nu}\partial _{\nu}$,
($g=|\det g_{\mu\nu}|$), is the D'Alambertian on a background manifold
$M$, and
$m$ is the mass of the field.
In general the integration contour in (\ref{eq:log}) should be
taken in complex pane. The parameter $\xi$ determines the coupling between
scalar field and curvature $R$ of $M$, and the case $\xi = 1/6$
corresponds to conformally invariant theory.
The kernel of the heat operator  $K_{M}(s)$
is the solution of the problem
\begin{equation}
\left(\partial/\partial s -\Box _{x}+ \xi R(x) \right)K_{M}(x,x's)=0~~~,
\label{eq:hk}
\end{equation}
\begin{equation}
K_{M}(x,x',0)=\delta _{M}(x,x')~~~
\label{eq:df}
\end{equation}
where $\delta _{M}(x,x')$ is the delta function covariantly defined on $M$.
For a smooth $d$ dimensional manifold without boundary
the asymptotic expansion of this kernel in powers of
the proper-time parameter $s$
looks as follows \cite{a2}
\begin{equation}
K_{M}(x,x',s)|_{s\rightarrow 0}={ e^{-\sigma ^2(x,x') /4s} \over (4\pi
s)^{d/2}}
\triangle^{1/2}(x,x')\sum_{n=0}^{\infty}a_n(x,x')s^n~~~.
\label{eq:smex}
\end{equation}
Here $\sigma(x,x')$ is
the geodesic distance between points $x$, $x'$ and $\triangle(x,x')$
is the Van Vleck determinant
\begin{equation}
\triangle(x,x')=-[g(x)g(x')]^{-1/2}\det\left({\partial ^2
\sigma(x,x')/2 \over \partial x^{\mu} \partial x'^{\nu}}\right)~~~.
\label{eq:biscalar}
\end{equation}
The coefficients $a_n$ of this expansion can be found
from the recursion relations and expressed in
terms of powers of the curvature tensor and its covariant derivatives
\cite{a2}.
Beside this, the first three coefficients are known to
define the structure of the one-loop counterterms in quantum theory
on $M$ \cite{a16}.

Consider a compact $d$-dimensional
manifold $M_{\alpha}$ having one-parameter group of isometry with the
Killing parameter $\varphi$ ranging from 0 to $\alpha$.
$M_{\alpha}$ can have Lorentzian or Euclidean signature.
Let the set $\Sigma$ of fixed points of the isometry group be
(n-2)-dimensional hypersurface near that $M_{\alpha}$ looks as a
space product $C_{\alpha}\times \Sigma$.
$C_{\alpha}$ is a conical
space with metric $ds^2=dr^2+r^2d\varphi^2$ and polar angle $\varphi$,
$0\leq\varphi \leq \alpha$.
If $\alpha\neq 2\pi$, $\Sigma$ is a singular surface where the scalar
curvature of $M_{\alpha}$ acquires a delta-function-like contribution.
On the other hand, outside this surface, no matter how close,
$M_{\alpha}$ has a smooth
geometry the same as in the case $\alpha = 2\pi$ when conical defects vanish.
We will use this fact to define the curvature tensor \footnote{Our convention
for the curvature
and Ricci tensors is $R^{\sigma}~_{\mu\nu\lambda}=\Gamma
^{\sigma}_{\mu\lambda,\nu}-
...$, and $R_{\mu\nu}=R^{\sigma}~_{\mu\sigma\nu}$.}
of $M_{\alpha}$ on $\Sigma$ and other geometrical quantities in terms of
their values for a smooth manifold $M\equiv M_{\alpha=2\pi}$. In particular,
we can introduce in this way orthogonal vectors to $\Sigma$ and describe
its imbedding in $M_{\alpha}$ by the same Gauss-Codacci equations as
imbedding in $M$. For convenience $\Sigma$ will be assumed to have two
orthogonal vectors $n_i$, $i=1,2$, normalized as $n^{\mu}_in_{j\mu}=
\delta _{ij}$.

Denote $K_{M_{\alpha}}$ the heat operator that is solution of
the problem (\ref{eq:hk}), (\ref{eq:df}) on $M_{\alpha}$, including
non-minimal coupling with the scalar curvature $R$.
Our aim of is to demonstrate that asymptotic
expansion of the trace of $K_{M_{\alpha}}$
has the structure similar to (\ref{eq:smex}) but
depends both on the conical angle $\alpha$
and on the local geometry near $\Sigma$
\begin{equation}
Tr~K_{M_{\alpha}}(s)|_{s\rightarrow 0}=
{1 \over (4\pi s)^{d/2}} \sum_{n=0}^{\infty}\left(a_n + a_{\alpha,n} \right)
s^n~~~.
\label{eq:singex}
\end{equation}
Here the standard coefficients $a_n=\int_{M_{\alpha}}a_n(x,x)d\Omega(x)$,
given by the integrals over the smooth domain of $M_{\alpha}$
($d\Omega(x)=\sqrt{g}d^nx$), get
modified, except $a_0$, by the surface terms
$a_{\alpha,n}$. The first three of these terms read
\begin{equation}
a_{\alpha,0}=0~~~,
{}~~~a_{\alpha,1}=\alpha c_1(\alpha)\int_{\Sigma}d\mu(\theta)~~~,
\label{eq:coeff''}
\end{equation}
$$
a_{\alpha,2}=
\int_{\Sigma}d\mu(\theta)\left[\alpha c_1(\alpha)
\left((\frac 16 -\xi)R(\theta) + \frac 16
R_{\mu\nu}(\theta)n^{\mu}_i n^{\nu}_i -
\frac 13 R_{\mu\nu\lambda\rho}(\theta)n^{\mu}_in^{\lambda}_in^{\nu}_jn^{\rho}_j
\right)+\right.
$$
\begin{equation}
\left.\alpha c_2(\alpha)\left(\frac 12
R_{\mu\nu\lambda\rho}(\theta)n^{\mu}_in^{\lambda}_in^{\nu}_jn^{\rho}_j-
\frac 14 R_{\mu\nu}(\theta)n^{\mu}_i n^{\nu}_i \right)\right]
\label{eq:coeff}
\end{equation}
where $\theta ^{\beta}$, $1\leq\beta\leq n-2$,
and $d\mu$ are coordinates and invariant volume on $\Sigma$, and
the curvature tensor is defined as it was explained above.
We also adopt the convention of summing Latin
indices $1\leq i,j \leq 2$. Finally, the numerical
coefficients $c_k(\alpha)$ (their definition follows below)
are polinomials in inverse even powers of $\alpha$, vanishing at
$\alpha =2\pi$.

The first term in (\ref{eq:singex}) is proportional
to the volume of $M_{\alpha}$ ($a_0=\int_{M_{\alpha}} d\Omega$) and
looks the same as for
the smooth manifolds. The second surface coefficient $a_{\alpha,1}$
was computed before for $M_{\alpha}=C_{\alpha}$ \cite{cheeger}
(see also \cite{b5}, \cite{ckv}). The next one, $a_{\alpha,2}$,
was also known for some particular cases. For $\xi=1/6$ it contributes
to the integral of conformal anomaly of the renormalized stress tensor
\cite{a6}, \cite{ckv}. So far as the scalar $R$ in (\ref{eq:coeff})
coincides with the curvature of the smooth manifold $M$,
we can consider imbedding of $\Sigma$ in $M$ and use the Gauss-Codacci
equations
\cite{schout} to represent $R$ in the form
\begin{equation}
R=R_{\Sigma}+2R_{\mu\nu}n^{\mu}_in^{\nu}_i-
R_{\mu\nu\lambda\rho}n^{\mu}_in^{\lambda}_in^{\nu}_jn^{\rho}_j
-(\chi _{i\mu}^{\mu})^2+\chi _i^{\mu\nu}\chi _{i\mu\nu}
\label{eq:gauss}
\end{equation}
where $R_{\Sigma}$ is the curvature of $\Sigma$ and $\chi ^i_{\mu\nu}$
are its second fundamental forms. Thus, for 2-dimensional surface
$a_{\alpha, 2}$ includes the Euler number of $\Sigma$.

All other quantities $a_{\alpha,n}$
have the structure similar to $a_{\alpha,2}$, eq. (\ref{eq:coeff}),
and depend on the conical angle deficit and curvature near
$\Sigma$.

\section{The heat kernel expansion}

It is worth reminding for further analysis the properties of the heat kernel
operator
$K_{C_{\alpha}}(r,r',\triangle \varphi,s)$,
$~\triangle\varphi $$=\varphi -\varphi '$,
on the conical space $C_{\alpha}$
that have been investigated in many papers \cite{cheeger},\cite{a5},\cite{a12}.
It can be expressed through the plane heat kernel
$K_{R^2}(r,r',\triangle \varphi, s)$ $=(4 \pi s)^{-1}\exp(-(x-x')^2/4s)$
by means of the Sommerfeld integral representation
\cite{a5},\cite{a12}
\begin{equation}
K_{C_{\alpha}}\left(r,r',\triangle\varphi,s\right)={i
\over 2\alpha} \int_{C} \cot \left(\pi \alpha^{-1} w\right)
K_{R^2}\left(r,r',\triangle\varphi +w,s\right) dw~~~,
\label{2.2}
\end{equation}
The integration
contour $C$ in (\ref{2.2}) has two branches, one in the upper half complex
plane of the parameter $w$ going from $(-\pi -\triangle\varphi+i\infty)$ to
$(\pi-\triangle\varphi +i\infty)$ and the other in the lower half-plane
from $(\pi-\triangle\varphi-i\infty)$ to $(-\pi-\triangle\varphi-i\infty)$,
see \cite{a12}. By using the above representation
the diagonal part of $K_{C_{\alpha}}$ can be shown to be a distribution
singular at the cone apex \cite{b5}, \cite{ckv}, whereas outside
this point it has the same asymptotic behaviour as the plane kernel $K_{R^2}$.
This fact indicates, in particular, that at $s\rightarrow 0$ conical
singularities give a nontrivial contributions only into trace of the
heat operators.

Consider now the manifold $M_{\alpha}$. It is convenient to begin
with the case $\alpha =2\pi$ when conical singularities are absent
and $\Sigma$ is the fixed-point set of $O(2)$ isometry group.
Take a point $p_0$ on $\Sigma$ and
introduce the Riemann normal coordinates $x^{\mu}$ with the origin at $p_0$.
In fact, we are interested in an interplay of two limits: when
$s\rightarrow +0$, and when a point $p(x)$, being in the smooth region,
approaches
$p_0$ ($x^{\mu}\rightarrow 0$). For this reason and to investigate
the integral properties of the heat kernel near $p_0$, it is
convenient to use the rescaled normal coordinates $y^{\mu}=s^{-1/2}x^{\mu}$.
Then at finite $y^{\mu}$ the small values of the parameter $s$ imply
that  $p(x)\rightarrow p_0$.

In terms of the rescaled normal
coordinates $y^{\mu}$ the standard expansion (\ref{eq:smex}) on
$M$ looks as follows
\begin{equation}
K_{M}(\sqrt{s}y,\sqrt{s}y',s)|_{p_0}={e^{-(y-y')^2/4} \over (4\pi s)^{d/2}}
\left(1+\sum_{n=0}^{\infty}b_{(n+2)/2}(y,y',p_0)s^{(n+2)/2}\right)
\label{eq:exx0}
\end{equation}
and holds near $p_0$ at $s\rightarrow 0$. Here $(y-y')^2=(y-y')^{\mu}
(y-y')_{\mu}$ and
 the leading term in
(\ref{eq:exx0})
is the plane kernel $K_{R^d}(\sqrt{s}y,$$\sqrt{s}y',s)$,
whereas the other terms represent the
corrections due to the curvature of the space $M$ at $p_0$. The
expressions of $b_{(n+2)/2}$ can be
obtained from (\ref{eq:smex}) after an additional decomposition of all
the quantities near $p_0$ in power series in $\sqrt{s}y^{\mu}$, $\sqrt{s}{y'}
^{\mu}$. For instance, by using the known asymptotic formulas \cite{donnelly1}
\begin{equation}
\triangle ^{1/2}(\sqrt{s}y,\sqrt{s}y')=1+{s \over 12}R_{\mu\nu}(p_0)
(y-y')^{\mu}(y-y')^{\nu}+O(s^{3/2})~~~,
\label{eq:as1}
\end{equation}
\begin{equation}
\sigma ^2(\sqrt{s}y,\sqrt{s}y')=s\left((y-y')^2-\frac s3 R_{\mu\lambda\nu
\rho}(p_0)(y-y')^{\mu}(y-y')^{\nu}y^{\lambda}y^{\rho}+
O(s^{3/2})\right)~~~,
\label{eq:as2}
\end{equation}
we obtain for $b_1$ the expression
\begin{equation}
b_1(y,y',p_0)=a_1(p_0)+{1 \over 12}\left(R_{\mu\nu}(p_0)(y-y')^{\mu}
(y-y')^{\nu}+R_{\mu\lambda\nu\rho}(p_0)(y-y')^{\mu}(y-y')^{\nu}y^{\lambda}
y^{\rho}\right)
\label{eq:b0}
\end{equation}
where $a_1(p_0)=a_1(x,x)|_{p(x)=p_0}= (1/6-\xi)R(p_0)$. On the other hand,
the next coefficient comes from the Taylor series of $a_1(x,x')$
\begin{equation}
b_{3/2}(y,y',p_0)=a_1,_{\mu}(p_0)(y'+y)^{\mu}~~~.
\label{eq:b1}
\end{equation}
It can be shown from the decompositions of $a_n(x,x')$, $\triangle ^{1/2}$ and
$\sigma ^2$ in Riemann coordinates series that all other coefficients
$b_{n/2}$ in (\ref{eq:exx0}) are expressed similar to (\ref{eq:b0}),
(\ref{eq:b1}) in terms of the curvature
tensor and its covariant derivatives.

Choose an orthonormal basis at $p_0$ in such a way to separate
the Riemanniain coordinates $y^{\mu}$ into $u^i$ coordinates,
$i=1,2$, associated to a basis orthogonal to $\Sigma$, and $v^{\beta}$,
$\beta =1,...,d-2$, corresponding to a tangent basis. The
coordinates $u^{i}$ can be parameterized by the polar angle
$0\leq \varphi \leq 2\pi$ and radius $u>0$ as usually,
$(u^1,u^2)=(u\cos\varphi, u\sin\varphi)$. Then $O(2)$ isometry
of $M$ generates the rotations $\varphi\rightarrow\varphi + \delta \varphi$
leaving unchanged $u$ and $v^{\beta}$.
Thus, we can use  mixed coordinates $y^{\mu}(\varphi)
\equiv (\varphi,u,v^{\beta})$ also for the singular space
$M_{\alpha}$. In these coordinates $M_{\alpha}$ is described
by the same metric but has different period of $\varphi$.
Besides, if $v^{\beta}=0$, the radius $\sqrt{s}u$ measures the distance to the
fixed point $p_0$.

To find for the singular heat operator $K_{M_{\alpha}}$ the
expansion similar to (\ref{eq:exx0}), introduce the following
{\it asymptotic} Sommerfeld-like representation
\begin{equation}
K_{M_{\alpha}}\left(\sqrt{s}y(\varphi),\sqrt{s}y'(\varphi'),s\right)|_{p_0}=
{i\over 2\alpha} \int_{C} \cot \left(\pi \alpha^{-1} w\right)
K_{M}\left(\sqrt{s}y(\varphi +w),\sqrt{s}y'(\varphi'),s\right)|_{p_0} dw~~~
\label{eq:asint}
\end{equation}
valid near $p_0$ at $s\rightarrow +0$
\footnote{Note, that due to the rotation symmetry $K_{M_{\alpha}}$ really
depends on the difference $\triangle\varphi =\varphi-\varphi '$.}.
Here the integration contour $C$ is chosen to be the same as in (\ref{2.2}) and
in the integrand the kernel $K_{M}$ should be considered
only in sense of the series (\ref{eq:exx0}). In this case the integral is
convergent for the each particular term of the sum due to the exponential
factor
$\exp(-uu'\cos(\triangle\varphi +w)/2)$ appearing from (\ref{eq:exx0}).
Other properties of this representation (\ref{eq:asint}) that enable us to
identify it near $p_0$ to the heat kernel operator on $M_{\alpha}$ are
as follows:

1) For each its angle argument $K_{M_{\alpha}}$, determined by
(\ref{eq:asint}),
is a function with period $\alpha$
\begin{equation}
K_{M_{\alpha}}(\sqrt{s}y(\varphi +\alpha),\sqrt{s}y'(\varphi '),s)|_{p_0}=
K_{M_{\alpha}}(\sqrt{s}y(\varphi),\sqrt{s}y'(\varphi '),s)|_{p_0}~~~.
\label{eq:periodicity}
\end{equation}

2) In neighborhood of $p_0$ the kernel (\ref{eq:asint}) satisfies
the heat equation (\ref{eq:hk}), written in the
mixed coordinates $(\varphi, \sqrt{s}u, \sqrt{s}v^{\beta})$, and to
"initial" condition
\begin{equation}
\lim_{s\rightarrow 0}\left[s^{d/2}\left(K_{M_{\alpha}}(\sqrt{s}y,
\sqrt{s}y',s)|_{p_0}-
K_{C_{\alpha}\times R^{d-2}}(\sqrt{s}y,\sqrt{s}y',s)\right)\right]=0
\label{eq:incond}
\end{equation}
analogous to the leading asymptotic of the smooth kernel $K_{M}$,
see (\ref{eq:exx0}). The meaning of this condition is that $K_{M_{\alpha}}$
should be locally represented as the kernel $K_{C_{\alpha}\times R^{d-2}}$
plus corrections due to the space curvature near $p_0$.

The first of these statements is a straightforward consequence of the
definition of the integral (\ref{eq:asint}). The second one holds so far as in
($\varphi,u,v^{\beta}$)
coordinates the operators on $M$ and $M_{\alpha}$ are described by the same
heat kernel equation (\ref{eq:hk}) that does not depend
on the angle $\varphi$.

Due to the local equivalence of $M_{\alpha}$ and $M$ outside
$\Sigma$ both kernels, $K_{M_{\alpha}}$,$K_{M}$, have
equal asymptotic expansions in the smooth region. This is
because the corresponding coefficients are determined by the same
local recursion relations. Therefore, to investigate the expansion
of $TrK_{M_{\alpha}}$ one can change $K_{M_{\alpha}}$ outside $\Sigma$
to $K_{M}$ and represent its trace in the form
$$
Tr~K_{M_{\alpha}}=\int_{M_{\alpha}}\left(K_{M_{\alpha}}(x,x,s)
-K_{M}(x,x,s)\right)d\Omega(x)+\int_{M_{\alpha}}K_{M}(x,x,s)d\Omega(x)=
$$
\begin{equation}
=\int_{\Sigma _{\epsilon}}\left(K_{M_{\alpha}}(x,x,s)-K_{M}(x,x,s)\right)
d\Omega(x)+\int_{M_{\alpha}}K_{M}(x,x,s)d\Omega(x)~+~ES
\label{eq:trace}
\end{equation}
valid at $s\rightarrow +0$ up to exponentially small terms $ES$.
Here $\Sigma _{\epsilon}$ is a sufficiently
small neighborhood of $\Sigma$ and $\epsilon$ is a small parameter associated
to its "thickness". So far as the diagonal part of $K_{M}$ does not depend
on $\varphi$, the last term in (\ref{eq:trace}) can be written as
$(\alpha /2\pi)TrK_{M}(s)$, for which the standard expansion is
applicable. On the other hand,
for the integral over domain $\Sigma _{\epsilon}$ we can use the
asymptotic formula (\ref{eq:asint}) valid near each of the point on
$\Sigma$.

By transforming in (\ref{eq:asint}) contour $C$ in the complex plane
we can write for the difference of the diagonal elements in $\Sigma
_{\epsilon}$
\begin{equation}
K_{M_{\alpha}}(\sqrt{s}y,\sqrt{s}y,s)|_{p_0}-K_{M}(\sqrt{s}y,\sqrt{s}y,s)|_{p_0}
={i \over 2\alpha}\int_{\Gamma}\cot(\pi\alpha ^{-1}w)
K_M(\sqrt{s}y(w),\sqrt{s}y,s)|_{p_0}dw
\label{eq:diff}
\end{equation}
where $\Gamma$ consists of two curves, going from $(-\pi+i\infty)$ to
$(-\pi -i\infty)$ and from $(\pi -i\infty)$ to $(\pi + i\infty)$ and
intersecting the real axis between the poles of the integrand
$-\alpha$, 0 and 0, $\alpha$ respectively.

Note, that for a point $p$ being sufficiently close to $\Sigma$
there is a unique geodesic line, starting from $p$ and orthogonal to $\Sigma$
at some point $p_0$. It means that $p$ can be completely determined by
$d-2$ coordinates $\theta ^{\beta}$ of $p_0$ on $\Sigma$,
by the geodesic distance $\sigma(p,p_0)$ and a polar angle $\varphi$.
On $M$ the rescaled Riemann coordinates of $p$ with the
origin at $p_0$ read as $y^{\mu}=(u\sin\varphi,$
$u\cos\varphi ,v^{\beta}=0)$ where $\sqrt{s}u=\sigma(p,p_0)$.
In this case (\ref{eq:exx0}) results to
\begin{equation}
K_{M}(\sqrt{s}y,\sqrt{s}y(w),s)|_{p_0(\theta)}=
{e^{-u^2\sin^2(w/2)} \over
(4 \pi s)^{d/2}}\sum_{n=0}^{\infty}b_{n}(u^2,w,\theta)s^n
\label{eq:as3}
\end{equation}
where $b_0\equiv 1$, $b_n(u^2,w,\theta)\equiv b_n(y,y(w),p_0(\theta)),~n\geq
1$.
Remarkably, all half integer
powers of $s$ in (\ref{eq:as3}) disappear due to the isometry.
Indeed, all the coefficients in this expansion are expressed
through the components of the Riemann tensor and its covariant derivatives
taken at the fixed point of $O(2)$ isometry group. Besides, each
power of $\sqrt{s}$ is associated to a Latin indices $i=1,2$ of these
quantities, that corresponds to $O(2)$ tensor representation, whereas tangent
Riemann coordinates, $v^{\beta}$, are absent. Thus,
so far as only even rank $O(2)$ invariant tensors are possible, we obtain
the integer power expansion (\ref{eq:as3}).
In particular, by using similar considerations,
$b_1$ coefficient in (\ref{eq:as3}) can be shown to read
\begin{equation}
b_1(u^2,w,\theta)=(\frac 16-\xi)R(\theta)+{1 \over 6}u^2\sin^2(w/2)
R_{ii}(\theta)
+{1 \over 24}u^4\sin^2w R_{ijij}(\theta)~~~.
\label{eq:b0'}
\end{equation}
Other $b_n(u^2,w,\theta)$, $n>1$, are represented, like (\ref{eq:b0'}), as
polinomials in invariant products $u^2$ and $(u-u(w))^2$ where the
maximal power of $(u-u(w))^2$, appearing from the decomposition (\ref{eq:as2})
of $\sigma ^2$, is equal to $n+1$.

Note that the integration in $\Sigma _{\epsilon}$ can be also
represented in terms of curvature corrections to the volume of the
space product $C_{\alpha}^{\epsilon}\times\Sigma$ with $\epsilon$ considered
as a cone radius
\begin{equation}
\int_{\Sigma_{\epsilon}}d\Omega(x)=\int_{\Sigma}d\mu(\theta)
\int_{0}^{\alpha}d\varphi \int_{0}^{\epsilon /\sqrt{s}}udu~s
\sum_{n=0}^{\infty}d_n(\theta)u^{2n}s^n~~~,
\label{eq:measure}
\end{equation}
\begin{equation}
d_0=1~~~,~~~
d_1={1 \over 6}R_{ijij}(\theta) -
\frac 14 R_{ii}(\theta)~~~.
\label{eq:meascoef}
\end{equation}
This decomposition and the form of $d_1$ for general case can be found
in \cite{donnelly1}.
Substituting (\ref{eq:as3}) in (\ref{eq:diff})
and integrating this expression in $\Sigma _{\epsilon}$
with the help of representatin (\ref{eq:measure}) we obtain
$$
\int_{\Sigma_{\epsilon}}d\Omega(x)\left[K_{M_{\alpha}}(x,x,s)-
K_{M}(x,x,s)\right]|_{s\rightarrow 0}=
$$
$$
{1 \over (4\pi s)^{d/2}}\sum_{n=0}^{\infty}\left(\int_{\Sigma}d\mu(\theta)
\frac i2 \int_{\Gamma}\cot(\pi\alpha^{-1}w)dw\int_{0}^{\epsilon/\sqrt{s}}
udue^{-u^2\sin^2w/2}\times \right.
$$
\begin{equation}
\left.
\sum_{m=0}^{n}d_m(\theta)b_{n-m}(u^2,w,\theta)u^{2m}
\right)s^{n+1}
\equiv{1 \over (4\pi s)^{d/2}}
\sum_{n=0}^{\infty}a_{\alpha,n+1}s^{n+1}+ES~~~.
\label{eq:diff2}
\end{equation}
Not difficult analysis based on the symmetry of the problem
shows that after integrating over $u^2$ in (\ref{eq:diff2})
$a_{\alpha,n}$ coefficients can be represented by series
\begin{equation}
a_{\alpha,n}=\sum_{k=1}^{n}\alpha c_k({\alpha})G_{nk}~~~.
\label{eq:alpha}
\end{equation}
The quantities $G_{nk}$ are expressed through the integrals
over $\Sigma$ on the powers of the curvature tensor and its covariant
derivatives and thus accumulate the information about local geometry near the
singular surface. On the other hand, global properties of $M_{\alpha}$
are contained in the integrals
\begin{equation}
c_n(\alpha)={i \over 4\alpha}\int_{\Gamma}\cot(\pi w\alpha^{-1})
(\sin w/2)^{-2n} dw
\label{3.7}
\end{equation}
that can be computed in terms of polynomials of the order $2n$ in
powers of $\alpha^{-1}$ \cite{a13}. In particular,
$c_1(\alpha),c_2(\alpha)$ in (\ref{eq:coeff}) can be written as
\begin{equation}
c_1(\alpha)=\frac 16 \left(\left(2\pi \alpha^{-1}\right)^2-1\right)~~,~~
c_2(\alpha)={1 \over 15} c_1(\alpha)
\left(\left(2\pi \alpha^{-1}\right)^2+11\right)~~~.
\label{3.10}
\end{equation}
Besides, the explicit form (\ref{eq:coeff}) of $a_{\alpha,1},~a_{\alpha,2}$
immediately follows from expressions (\ref{eq:b0'}) for $b_1$ and
(\ref{eq:meascoef}) for $d_1$. Finally, the asymptotic expansion
(\ref{eq:singex}) for the trace of the heat kernel operator on
$M_{\alpha}$ can be
obtained by substituting (\ref{eq:smex}), (\ref{eq:diff2}) in (\ref{eq:trace}).

The general results (\ref{eq:singex}) - (\ref{eq:coeff}) for
singular heat kernel expansion and first coefficients coincide
with expressions previously obtained for some particular
cases.

The first example of this
kind can be found in works of H.Donnelly \cite{donnelly1}, \cite{donnelly2}
where asymptotic formulas for the trace on quotients of compact manifolds
with fixed-point isometry groups have been studied.
His analysis can be readily used to evaluate both general form and
first three heat coefficients (see theorems 4.1, 5.1 in \cite{donnelly1})
for cyclic rotation groups, which corresponds
to singularities with conical angle taking discrete values $\alpha=2\pi n^{-1},
{}~(n=2,3,...)$. It can be shown that these results are in agreement with
the ones derived here by the different method for arbitrary values of
angle $\alpha$.

The other interesting example \cite{a6} to be mentioned here concerns
computation of zeta-function $\zeta(z)$ on a spherical domain
described by the 4-dimensional spherical metric
$ds^2=d\theta _3^2+\sin^2\theta _3d\theta _2^2$$+\sin^2\theta _3\sin ^2\theta
_2
d\theta _1^2+\sin^2\theta _3\sin ^2\theta _2\sin^2\theta _1d\varphi ^2$ but
with arbitrary ranging of the polar angle $0\leq\varphi\leq\alpha$.
The value of $\zeta(0)$, obtained explicitly in \cite{a6}, by using the
the spectrum of the Laplace operator on this space, can be also rederived
on the base of formulas (\ref{eq:singex}) - (\ref{eq:coeff}) with the help
of the Mellin representation for $\zeta(z)$.

\section{Conclusions and remarks}

In this letter the general structure of the heat kernel expansion
and explicit expression for the first three coefficients was found for
manifolds with conical singularities of the form
$C_{\alpha}\times\Sigma$.
These results were shown to be confirmed by calculations
in particular cases.

For quantum field theory conical singularities qualitatively result to
similar consequences as in presence of boundaries of the background
spaces \cite{b5}. For instance, in this theory the renormalized stress
tensor diverges in non-integrable way as a singular point is approached
\cite{a17},\cite{a13}, whereas the one-loop effective action and
other integral quantities acquire additional surface ultraviolet divergent
terms. However, as compared to case with boundaries only integer
powers of the proper time parameter $s$ are present in the heat kernel
expansion (\ref{eq:singex}).

In classical theory conical singularities are associated to a matter
distribution with a surface energy determined by the angle deficit
$2\pi-\alpha$. For the flat conical geometry the only new divergences arise
from the coefficient $a_{\alpha,1}$ in (\ref{eq:singex}). They contribute
to the classical surface matter action and can be, consequently, removed by
renormalization of $\alpha$. In quantum field theory around cosmic strings
it is equivalent to a renormalization of the string tension \cite{b5}.
However, in general this renormalization is not sufficient and some
additional surface counterterms should be introduced to get rid off
the divergences given by the next coefficient $a_{\alpha,2}$.
According to (\ref{eq:coeff}) they are determined by the space
curvature and have a more complicated structure than initial classical action
that is simply proportional to the area of $\Sigma$. It means that when conical
singularities are present one should allow to the generalized
Einstein action \cite{a16} an addition in the form of an effective surface
functional with terms having the structure of $a_{\alpha,2}$.
In particular, the relevance of nontrivial surface terms in the effective
gravitational action has been recently discussed in \cite{b4} for the
black hole thermodynamics.

Finally, it is worth mentioning that the technique based on the asymptotic
Sommerfeld-like representation for the scalar heat kernel operator can also be
developed for higher spins, which has a number of implications.

I would like to thank Professor W. Drechsler for helpful discussions
and hospitality during my visit in the Max-Planck-Institute f{\"u}r
Physik in M{\"u}nchen. This work was supported in part by the International
Science Foundation (Soros) Grant No. Ph1-0802-0920.

\end{document}